\documentclass[11pt,aps,nofootinbib]{revtex4-1}
\usepackage{amsfonts}
\usepackage{mathrsfs}
\usepackage{amsmath}
\usepackage{amssymb}
\usepackage[normalem]{ulem}
\usepackage{extarrows}
\usepackage{slashed}
\usepackage{isodateo}
\usepackage{graphicx} 
\usepackage{xcolor}
\usepackage[bookmarksnumbered=true,bookmarksopen=true]{hyperref}
 \hypersetup{colorlinks,%
             linkcolor=[rgb]{0,0.3,0.6}, %
             citecolor=[rgb]{0,0.3,0.6}, %
             urlcolor=[rgb]{0,0.3,0.6}}%
\renewcommand{\rm}{\mathrm} 
\begin{document}

\title{Minimal Vectorlike Model in Supersymmetric Unification}
\author{Sibo Zheng$^{1,2,}$}
\email{sibozheng.zju@gmail.com}
\affiliation{$^{1}$Department of Physics, Chongqing University, Chongqing 401331, China \\
$^{2}$Department of Physics, Harvard University, Cambridge, MA 02138, USA}

\begin{abstract}
Compared to the minimal supersymmetric standard model,
an extension by vectorlike fermions is able to explain the Higgs mass while retains the grand unification.
We investigate the minimal vectorlike model by focusing on the vectorlike electroweak sector.
We firstly derive the mass spectrum in the electroweak sector,
then calculate the one-loop effects on the Higgs physics,
and finally explore either vectorlike or neutralino dark matter.
Collider constraints are briefly discussed. 
\end{abstract}

\maketitle
\section{Introduction}
With the discovery of Higgs scalar \cite{Higgsmass1,Higgsmass2} at the Large Hadron Collider (LHC), 
identifying the nature of this scalar, 
which reveals the origin of electroweak symmetry breaking (EWSB), 
is one of prior tasks at this facility.
There are various scenarios in the literature that naturally explain the EWSB,
among which supersymmetry (SUSY) has attracted much attention since the discovery.
Nevertheless, the prospective on the  minimal supersymmetric standard model (MSSM) is rather pessimistic.
Some reasons for this include that the mass parameters related to the third-generation squarks 
are probably at least of order several TeVs \cite{MSSMHiggs1,MSSMHiggs2},
and also signals of neutralino dark matter aren't yet observed at any dark matter direct detection facilities \cite{MSSMDark1,MSSMDark2}.

The status of MSSM motivates a diversity of extensions,
of which vectorlike (VL) fermions are interesting because of a few features as follows.
\begin{itemize}
\item First, the VL fermions are one of viable extensions which retain the perturbative unification \cite{1711.05362,1706.01071}, 
similar to a well-known example - the next-to-minimal supersymmetric model (NMSSM).
\item Second, the issue of naturalness imposed by the Higgs mass in the MSSM can be resolved by new radiative correction to the Higgs mass 
due to the VL fermions of order TeV scale \cite{0910.2732, MO, Babu}.
It differs from the NMSSM, as in the later one the correction occurs at the tree level.
\item Last, the stringent constraint on the neutralino dark matter from dark matter direct detection will be modified due to be presence of neutral fermions in the VL sector.
\end{itemize}
This paper is devoted to study the minimal VL model as described in Table.\ref{Models}.
We will explore this model by focusing on the VL leptons therein,
the matter content of which is composed of one singlet and two doublets.
A key point in this model is that the VL leptons directly couple to the Higgs doublets rather than mix with the SM leptons,
which suggests that they may play important roles in both the Higgs physics and dark matter.

\begin{table}[htb!]
\begin{center}
\begin{tabular}{ccc}
\hline\hline
VL model    & ~~Representation    \\  \hline
Minimal   &$\mathbf{1}+\mathbf{5}+\mathbf{\bar{5}}$ \\
LND \cite{0910.2732}  &$\mathbf{1}+\mathbf{1}+\mathbf{5}+\mathbf{\bar{5}}$ \\
 VL  4th-gen \cite{Babu}  &~~$\mathbf{1}+\mathbf{5}+\mathbf{\bar{5}}+\mathbf{10}+\mathbf{\bar{10}}$\\
\hline \hline
\end{tabular}
\caption{TeV-scale VL models and their representations under ${\rm SU(5)}$.}
\label{Models}
\end{center}
\end{table}

The paper is organized as follows.
In Sec.II, we introduce the minimal VL model, 
derive the EWSB conditions, and calculate mass spectrum in the electroweak sector for our purpose.
Sec.III is devoted to estimate effects on the Higgs physics,
where the one-loop radiative correction to Higgs mass 
and the modification on Higgs self coupling are presented in Sec.IIIA and Sec.IIIB, respectively. 
Sec.IV is devoted to discuss the dark matter physics in the presence of VL leptons,
where in Sec.IVA we will show the VL dark matter is excluded by the large Yukawa coupling as required by the Higgs mass,
whereas in Sec.IVB the parameter space of neutralino dark matter is expanded due to the relax of Higgs mass constraint.
In Sec.V, we briefly discuss the collider constraints on the VL  leptons.
Finally, we conclude in Sec.VI.

\section{The Model}
In the minimal VL model, the matter content contains a singlet $N$, two down-type triplets $D$ and $\bar{D}$, 
and two VL  doublets $L$ and $\bar{L}$ with hyper charge $1/2$ and $-1/2$, respectively.
They constitute a $\mathbf{5}_{V}$ and a $\bar{\mathbf{5}}_{V}$ representation of SU(5)
consistent with grand unification \cite{1706.01071,1711.05362}.

The superpotential for the VL electroweak sector reads as,
 \begin{eqnarray}{\label{VLs}}
W_{VL}=kH_{u}\bar{L}N-hLH_{d}N+M_{L}\bar{L}L+\frac{1}{2}M_{N}N^{2}
\end{eqnarray}
where $M_{L,N}$ refer to VL masses and $k$ and $h$ are Yukawa coupling constants.
The superpotential for MSSM is given by,
 \begin{eqnarray}{\label{MSSMs}}
W_{\text{MSSM}}=Y_{u}q\bar{u}H_{u}+Y_{d}q\bar{d}H_{d}+Y_{e}l\bar{e}H_{d}+\mu H_{u}H_{d}.
\end{eqnarray}
Although the matter content in Eq.(\ref{VLs})  is similar to that of NMSSM with two additional doublets \cite{Fayet},
they are different as what follows. 
Following our purpose as mentioned above,
we assume that the conservative $R$-parities of $\mathbf{5}_{V}$ and $\bar{\mathbf{5}}_{V}$ that contain $L$ and $\bar{L}$ are the same as those of the $\mathbf{5}_{H}$ and $\bar{\mathbf{5}}_{H}$ that include the Higgs doublets $H_{u}$ and $H_{d}$.
Moreover, in order to ensure that the low-energy effective superpotential is precisely described by Eq.(\ref{VLs}) and Eq.(\ref{MSSMs}), 
we impose a $Z_2$ parity, 
under which $N$, $\mathbf{5}_{V}$ and $\bar{\mathbf{5}}_{V}$ referring to the VL sector are odd but $\bar{\mathbf{5}}_{M}$, $\mathbf{10}_{M}$, $\mathbf{5}_{H}$ and $\bar{\mathbf{5}}_{H}$ referring to the MSSM sector are even.
The $Z_2$ parity is spontaneously broken by the vacuum expectation values (vevs) in the VL sector.
Since the triplets $D$ and $\bar{D}$ affect the phenomenological analysis\footnote{
Neither they directly couple to the Higgs doublets nor mix with the neutralinos/VL electroweak fermions. Thus, both the corrections to the Higgs physics and dark matter analysis are expected to be small. 
Note, although perturbative unification favors VL triplet mass $M_{D}\sim M_L$, 
it may be still intact even for a moderate mass splitting between $M_{D}$ and $M_L$.
Because it is not only sensitive to the mass spectrum of VL sector but also that of MSSM sector, and even intermediate physics between the weak scale and unification scale.} only in terms of their interaction with heavy triplet fields in $\mathbf{5}_{H}$ and $\bar{\mathbf{5}}_{H}$, 
we will simply neglect them.

Apart from the superpotential, 
the soft mass Lagrangian $\mathcal{L}_{\text{soft}}$ is also expanded as,
 \begin{eqnarray}{\label{soft}}
-\mathcal{L}_{\text{soft}}&\supset&-\mathcal{L}^{\text{MSSM}}_{\text{soft}}
+m^{2}_{L}\mid L\mid^{2}+m^{2}_{\bar{L}}\mid \bar{L}\mid^{2}+m^{2}_{N}\mid N\mid^{2}\nonumber\\
&+&\left(kA_{k}H_{u}\bar{L}N-hA_{h}LH_{d}N+\text{H.c}\right),
\end{eqnarray}
where $m_{L,N,D}$ and $A_{k,h}$ represent soft mass parameters, 
and $\mathcal{L}^{\text{MSSM}}_{\text{soft}}$ contains MSSM soft masses such as
 \begin{eqnarray}{\label{Higgs}}
-\mathcal{L}^{\text{MSSM}}_{\text{soft}}= m^{2}_{H_{u}}\mid H_{u}\mid^{2}+m^{2}_{H_{d}}\mid H_{d}\mid^{2}+(bH_{u}H_{d}+\text{H.c}).
\nonumber
\end{eqnarray}

Following the notation and conventions in \cite{9709356}, 
we express the Higgs and VL doublets as, respectively,
\begin{eqnarray}{\label{Dec1}}
H_{u}&=&\left(%
\begin{array}{c}
H^{+}_{u}  \\
H^{0}_{u}\\
\end{array}%
\right)=\left(%
\begin{array}{c}
H^{+}_{u}  \\
\upsilon_{u}+\frac{1}{\sqrt{2}}(H^{0}_{uR}+iH^{0}_{uI})\\
\end{array}%
\right)\nonumber\\
H_{d}&=&\left(%
\begin{array}{c}
H^{0}_{d}  \\
H_{d}^{-}\\
\end{array}%
\right)=\left(%
\begin{array}{c}
\upsilon_{d}+\frac{1}{\sqrt{2}}(H^{0}_{dR}+iH^{0}_{dI})  \\
H^{-}_{d}\\
\end{array}%
\right)
\end{eqnarray}
and
\begin{eqnarray}{\label{Dec1}}
L&=&\left(%
\begin{array}{c}
E^{+}  \\
\eta\\
\end{array}%
\right)=\left(%
\begin{array}{c}
E^{+}  \\
v_{L}+\frac{1}{\sqrt{2}}(\eta_{R}+i\eta_{I})\\
\end{array}%
\right)\nonumber\\
\bar{L}&=&\left(%
\begin{array}{c}
\bar{\eta} \\
E^{-} \\
\end{array}%
\right)=\left(%
\begin{array}{c}
\bar{v}_{L}+\frac{1}{\sqrt{2}}(\bar{\eta}_{R}+i\bar{\eta}_{I})  \\
E^{-}\\
\end{array}%
\right)
\end{eqnarray}
Similarly, we write singlet $N$ as 
\begin{eqnarray}{\label{Dec2}}
N=n+\frac{1}{\sqrt{2}}(N_{R}+iN_{I}).
\end{eqnarray}
Mass parameters $\upsilon_{u,d}$, $v_{L}$, $\bar{v}_{L}$ and $n$ denote vevs.

These vevs are determined by the minimization of scalar potential $V$ as given by 
\begin{eqnarray}{\label{V}}
V=V_{F}+V_{D}-\mathcal{L}_{\text{soft}}
\end{eqnarray}
where
\begin{eqnarray}{\label{Vs}}
V_{F}&=&\mid \mu H_{d}+k\bar{L}N\mid^{2}+ \mid \mu H_{u}-hLN\mid^{2}\nonumber\\
&+& \mid M_{L}\bar{L}-hH_{d}N\mid^{2}+ \mid kH_{u}N+M_{L}L\mid^{2}
+\mid kH_{u}\bar{L}-hLH_{d}+M_{N}N\mid^{2},\nonumber\\
V_{D}&=&\frac{g^{2}_{1}+g^{2}_{2}}{8}(\mid H_{u}\mid^{2}-\mid H_{d}\mid^{2}+\mid L\mid^{2}-\mid \bar{L}\mid^{2})^{2}
+\frac{g^{2}_{2}}{2}\mid H_{u}H^{\dag}_{d}+L\bar{L}^{\dag}\mid^{2}.
\end{eqnarray}
Here, $g_{1}$ and $g_2$ refers to $U(1)_Y$ and  $SU(2)_{L}$ gauge coupling constant, respectively. 
Substituting Eq.(\ref{Dec1}) and Eq.(\ref{Dec2})  into Eq.(\ref{V}) gives rise to the EWSB conditions: 
\begin{widetext}
\begin{eqnarray}{\label{minimal}}
0&=& F_{N}(-k\bar{v}_{L})+  F_{\bar{L}} (kn)+F_{H_{d}}\mu-kA_{k}n\bar{v}_{L}+m^{2}_{H_{u}}\upsilon_{u}-\frac{1}{2}M^{2}_{Z}\cos2\beta\upsilon_{u}(1-\delta)-b\cot\beta \upsilon_{u}\nonumber\\
0&=& F_{N}(hv_{L})+  F_{L} (-hn)+F_{H_{u}}\mu+hA_{h}nv_{L}+ m^{2}_{H_{d}}\upsilon_{d}+\frac{1}{2} M^{2}_{Z}\cos2\beta\upsilon_{d}(1-\delta)-b\tan\beta\upsilon_{d} \nonumber\\
0&=&h^{2}n^{2}v_{L}+F_{N}(h\upsilon_{d})+hA_{h}n\upsilon_{d}+m^{2}_{L}v_{L}-\frac{1}{2}M^{2}_{Z}\cos2\gamma v_{L} \delta \nonumber\\
0&=& k^{2}n^{2}\bar{v}_{L}+F_{N}(-k\upsilon_{u})-kA_{k}n\upsilon_{u}+m^{2}_{\bar{L}}\bar{v}_{L}+\frac{1}{2}M^{2}_{Z}\cos2\gamma \bar{v}_{L} \delta \nonumber\\
0&=&F_{N}M_{N}+F_{L} (-h\upsilon_{d})+F_{\bar{L}}(k\upsilon_{u})+n(k^{2}\bar{v}^{2}_{L}+h^{2}v^{2}_{L})-kA_{k}\upsilon_{u}\bar{v}_{L}+hA_{h}\upsilon_{d}v_{L}+ m^{2}_{N}n,
\end{eqnarray}
\end{widetext}
where $F_i$ refer to the $F$-terms related to chiral superfield $i=\{N, L, \bar{L}, H_{u,d}\}$. 
For simplicity, we have neglected small cross terms from $D$-terms.
In Eq.(\ref{minimal}), $\tan\beta$, $\tan\gamma$ and $\delta$ are defined as 
\begin{eqnarray}{\label{defs}}
\tan\gamma&=&\frac{v_{L}}{\bar{v}_{L}}, ~~~~~~~~~\delta=\frac{v^{2}_{L}+\bar{v}^{2}_{L}}{(\upsilon/\sqrt{2})^{2}}, \nonumber\\
\tan\beta&=&\frac{\upsilon_{u}}{\upsilon_{d}},~~~~ 1-\delta=\frac{\upsilon^{2}_{u}+\upsilon^{2}_{d}}{(\upsilon/\sqrt{2})^{2}},
\end{eqnarray}
with the expectation value for the weak scale $\upsilon\simeq 246$ GeV.

With the conditions of EWSB described in Eq.(\ref{minimal}), 
one can directly derive the scalar and fermion mass spectrum in the electroweak sector.
We refer the reader to appendix A and B for scalar mass spectrum such as CP-even, CP-odd, and CP-charged scalars, 
and fermion mass spectrum such as neutralinos and charginos, respectively. 
In the next section, we will discuss constraints on the model parameters from precision measurements on the Higgs couplings.

\section{Higgs Physics}
\subsection{Higgs Couplings}
The precision measurement on the Higgs couplings is one of major tasks at the LHC,
for it reveals the pattern of EWSB in the sense that
different new physics models predict different sets of Higgs couplings.
The LHC has verified the SM-like Higgs coupling to SM fermions such as $b$ \cite{1709.07497} and  $\tau$ \cite{1501.04943}
as well as couplings to SM vector bosons such as $W$ \cite{1412.2641}.
These measurements tell us to what extension \cite{1303.3879,1606.02266,1809.10733} $\upsilon$ is deviated from its SM reference value. 
Such data will be further improved in higher level of precision at the future HL-LHC \cite{1310.8361} or ILC \cite{1310.0763} in preparation.

\begin{table}
\begin{small}
\begin{center}
\setlength{\tabcolsep}{0mm}{
\begin{tabular}{|c|c|c|}
\hline
 & \ \text{Parameter} &\ \text{Fit}  \\
\hline
$k_t$ &  $\frac{1}{\sqrt{1-\delta}}\frac{\mid\mathcal{O}_{11}\mid}{s_{\beta}}$   &  $0.81^{+0.19}_{-0.15}$  \\
$k_b$ &  $\frac{1}{\sqrt{1-\delta}}\frac{\mid\mathcal{O}_{12}\mid}{c_{\beta}}$   &  $0.74^{+0.33}_{-0.29}$  \\
$k_{\tau}$ &  $\frac{1}{\sqrt{1-\delta}}\frac{\mid\mathcal{O}_{21}\mid}{c_{\beta}}$   &  $0.84^{+0.19}_{-0.18}$  \\
$k_W$ &  $\mid\sqrt{1-\delta} (s_{\beta}\mathcal{O}_{11}+c_{\beta}\mathcal{O}_{12})+\sqrt{\delta} (s_{\gamma}\mathcal{O}_{13}+c_{\gamma}\mathcal{O}_{14})\mid$   &  $0.95^{+0.14}_{-0.13}$  \\
$k_Z$ &  $\mid\sqrt{1-\delta} (s_{\beta}\mathcal{O}_{11}+c_{\beta}\mathcal{O}_{12})+\sqrt{\delta} (s_{\gamma}\mathcal{O}_{13}+c_{\gamma}\mathcal{O}_{14})\mid$   &  $1.05^{+0.16}_{-0.16}$  \\
$\text{vev}$ &  $\upsilon\sqrt{1-\delta}$  & $231^{+13}_{-15}$\\
  \hline
\end{tabular}}
\caption{LHC constraints from global fits to SM-like Higgs couplings  \cite{1412.8662} and weak scale $\upsilon$ \cite{1809.10733} at $68\%$ CL. Mass parameter is in unit of GeV.}
\label{fits}
\end{center}
\end{small}
\end{table}

The Higgs couplings to SM particles in our model are given as,
\begin{small}
\begin{eqnarray}{\label{couplings}}
y_{t}&=& \frac{m_{t}}{\upsilon} \frac{\mathcal{O}_{11}}{\sqrt{1-\delta}s_{\beta}} , \nonumber\\
y_{b}&=& \frac{m_{b}}{\upsilon} \frac{\mathcal{O}_{12}}{\sqrt{1-\delta}c_{\beta}}, \nonumber\\
y_{\tau}&=&\frac{m_{\tau}}{\upsilon} \frac{\mathcal{O}_{12}}{\sqrt{1-\delta}c_{\beta}}  , \nonumber\\
y_{W}&=& \frac{2M^{2}_{W}}{\upsilon}\left[\sqrt{1-\delta}(s_{\beta}\mathcal{O}_{11}+c_{\beta}\mathcal{O}_{12})
+\sqrt{\delta}(s_{\gamma}\mathcal{O}_{13}+c_{\gamma}\mathcal{O}_{14})\right], \nonumber\\
y_{Z}&=& \frac{M^{2}_{Z}}{\upsilon}\left[\sqrt{1-\delta}(s_{\beta}\mathcal{O}_{11}+c_{\beta}\mathcal{O}_{12})
+\sqrt{\delta}(s_{\gamma}\mathcal{O}_{13}+c_{\gamma}\mathcal{O}_{14})\right],
\end{eqnarray}
\end{small}
where the constant coefficients are the SM values \footnote{The Higgs couplings $y_{W}$ and $y_Z$ receive contributions both from the expansions of the kinetic energy terms of Higgs doublets $H_{u,d}$ and VL doublets $L$ and $\bar{L}$, 
which are proportional to the vevs $\upsilon_{u,d}$ and $v_{L},\bar{v}_{L}$, respectively. 
Following the definitions in Eq.(\ref{defs}), we read the forms of $y_{W}$ and $y_{Z}$.}.
$\mathcal{O}_{ij}$ are components of the orthogonal matrix $\mathcal{O}$ which diagonalizes the CP-even mass matrix $\mathcal{M}^{2}_{S}$ in appendix A.
In terms of $\mathcal{O}_{ij}$ the SM-like Higgs scalar $h_1$ can be written as,
\begin{eqnarray}{\label{O}}
h_{1} =\mathcal{O}_{11} H^{0}_{uR} +\mathcal{O}_{12} H^{0}_{dR} + \mathcal{O}_{13} \eta_{R} +\mathcal{O}_{14} \bar{\eta}_{R} +\mathcal{O}_{15} N_{R}.
\end{eqnarray}

Unlike in the case of two Higgs doublets, 
where it is convenient to parameterize the couplings of SM-like Higgs scalar in term of two free angles $\beta$ and $\alpha$,
in our model there are five new parameters $m_{N}$, $m_{L,\bar{L}}$ and $A_{k,h}$ and three new EWSB conditions as shown in Eq.(\ref{minimal}).
It implies that in our case there are four free parameters related to the Higgs couplings,
which are subject to the constraints from the global fits to the LHC measurements on the Higgs couplings as summarized in Table.\ref{fits}.
In the following analysis, we choose the set of parameters $\tan\beta$, $\mathcal{O}_{11}$, $\mathcal{O}_{12}$ and $\delta$ for later discussions. 

The best fit value for $\delta$ has changed from the earlier $244$ GeV in ref.\cite{1303.3879} to $233$ GeV in ref.\cite{1606.02266} and $231$ GeV in ref.\cite{1809.10733}. 
The latest constraint on $\delta$ in Table.\ref{fits} suggests the following parameter range,
\begin{eqnarray}{\label{delta}}
0\leq \delta\leq 0.22,
\end{eqnarray}
which implies that the EWSB is dominated by the Higgs doublet vevs.
So, it is rational to ignore the small $\sqrt{\delta}$ term in the scaling factors $k_{W}$ and $k_{Z}$ in Table.\ref{fits}.
With these approximations,  we obtain the constraint on $\mathcal{O}_{11}$ and $\mathcal{O}_{12}$ in Table.\ref{fits}
\begin{eqnarray}{\label{ps}}
0.78 \leq \mathcal{O}^{2}_{11}+\mathcal{O}^{2}_{12}\leq 1.0,
\end{eqnarray}
after we perform a numerical scan within the region $10\leq\tan\beta\leq 50$.

\subsection{Higgs Self Coupling}
Apart from the precision measurements on the Higgs couplings above,
a probe of Higgs self interaction is also useful in identifying the nature of EWSB.
In the SM, the tri-Higgs and quartic Higgs coupling reads as, respectively,
\begin{eqnarray}{\label{selfsm}}
\lambda^{\text{SM}}_{3h}&=&\frac{3m^{2}_{h}}{\upsilon},\nonumber\\
\lambda^{\text{SM}}_{4h}&=&\frac{3m^{2}_{h}}{\upsilon^{2}},
\end{eqnarray}
where the self coupling $\lambda_{3h}$ is a key parameter to determine the SM Higgs pair production at LHC \cite{9603205,GB,9805244,1305.7340,1309.6594},
which mainly arises from gluon gluon fusion.
The cross section for Higgs pair production is altered in new physics such as MSSM and NMSSM even at tree level \cite{9603205,9805244, 1206.5816, 1304.3670,1305.6397,1504.06932}.

\begin{figure}
\centering
\includegraphics[width=8cm,height=8cm]{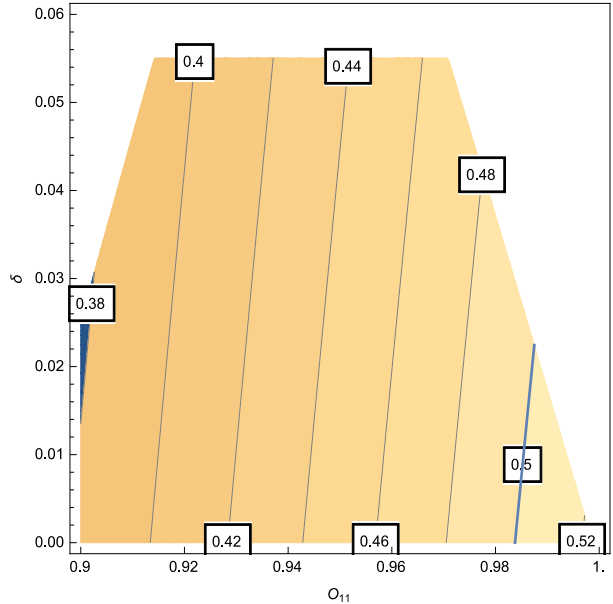}
\centering
\caption{The contour of $\lambda_{3h}/\lambda^{\text{SM}}_{3h}$ as function of $\delta$ and $\mathcal{O}_{11}$ for $\tan\beta=20$ and $\mathcal{O}_{12}=0.05$. 
The shaded region satisfies all constraints in the Table.\ref{fits},
where triangle region on the right side of green curve is in the reach of HL-LHC ($14$ TeV, 3$\text{ab}$$^{-1}$) \cite{1310.8361}. }
\label{lambda3h}
\end{figure}

In our case, both $\lambda_{3h}$ and $\lambda_{4h}$ receive contributions from $V_F$ and $V_D$, which are given by,
\begin{eqnarray}{\label{selfvl}}
\lambda_{3h_{1}}&=&\frac{3M^{2}_{Z}}{\upsilon}\left[\sqrt{1-\delta}(\mathcal{O}^{2}_{11}-\mathcal{O}^{2}_{12})(s_{\beta}\mathcal{O}_{11}-c_{\beta}\mathcal{O}_{12})+\sqrt{\delta}(\mathcal{O}^{2}_{13}-\mathcal{O}^{2}_{14})(s_{\gamma}\mathcal{O}_{13}-c_{\gamma}\mathcal{O}_{14})\right],\nonumber\\
\lambda_{4h_{1}}&=&\frac{3M^{2}_{Z}}{\upsilon^{2}}\left[(\mathcal{O}^{2}_{11}-\mathcal{O}^{2}_{12})^{2}+(\mathcal{O}^{2}_{13}-\mathcal{O}^{2}_{14})^{2}\right].
\end{eqnarray}
From Eq.(\ref{ps}), one observes that the components $\mathcal{O}_{11}$ and $\mathcal{O}_{12}$ 
constitute most of SM-like Higgs within region with moderate or large $\tan\beta$.
In contrast, the other components $\mathcal{O}_{1j}$ ($j=3-5$) can be ignored.
In Eq.(\ref{selfvl}), we have neglected contributions to $\lambda_{3h}$ and $\lambda_{4h}$ due to $V_F$ which are at least of order $\mathcal{O}^{2}_{1j}$.
As expected, the results in Eq.(\ref{selfvl}) reduce to those of MSSM under the limit $\delta\rightarrow 0$,
\begin{eqnarray}{\label{selfmssm}}
\lambda^{\text{MSSM}}_{3h_{1}}&=&\frac{3M^{2}_{Z}}{\upsilon}\cos(2\alpha)\sin(\alpha+\beta),\nonumber\\
\lambda^{\text{MSSM}}_{4h_{1}}&=&\frac{3M^{2}_{Z}}{\upsilon^{2}}\cos^{2}(2\alpha).
\end{eqnarray}  
The probe of Higgs self interaction is thus useful in discriminating this model from the others.

Following the constraints on $\mathcal{O}_{11}$ and $\mathcal{O}_{12}$ in the Table.\ref{fits},
we show in Fig.\ref{lambda3h} the contour of $\lambda_{3h}/\lambda^{\text{SM}}_{3h}$ for $\tan\beta=20$ and $\mathcal{O}_{12}=0.05$,
with the shaded region satisfying all constraints in the Table.\ref{fits}.
Note the ratio $\lambda_{3h}/\lambda^{\text{SM}}_{3h}$ is more sensitive to scaling factor $\delta$ and $\mathcal{O}_{11}$ rather than $\mathcal{O}_{12}$ for large $\tan\beta$.
Under the limit $\delta\rightarrow 0$, the ratio approaches to the maximal value $\lambda_{3h}^{\text{max}}/\lambda^{\text{SM}}_{3h}$ $\sim 0.52$.
With $\delta\neq 0$, a suppression on the ratio appears.
We find that the triangle region on the right side of green curve is in the reach of HL-LHC ($14$ TeV, 3$\text{ab}$$^{-1}$) \cite{1310.8361},
whereas the left side will be still invisible in the foreseeable future.

\subsection{Higgs Mass}
The soft mass parameters in Eq.(\ref{soft}) lead to mass splittings between the fermion and its scalar partner masses 
in VL superfield $L$, $\bar{L}$ and $N$.
It can be extracted via evaluating the one-loop correction to effective potential \cite{CW},
\begin{eqnarray}{\label{potential}}
\Delta V= 2\sum^{3}_{i=1} [F(M^{2}_{b_{i}})-F(M^{2}_{f_{i}})],
\end{eqnarray}
with $F(x)=x^{2}[\ln(x/Q^{2})-3/2]/64\pi^{2}$.
For simplicity, we consider the universal soft masses $m_{L}=m_{\bar{L}}=m_{N}=m$, universal VL masses $M_{L}=M_{\bar{L}}=M_{N}=M$ and large $\tan\beta$ limit.
Meanwhile, we take the good approximation $v_{L}=\bar{v}_{L}=0$.
The neutral scalar mass squared matrix, which determines the mass eigenvalues $M^{2}_{b_{i}}$ in Eq.(\ref{potential}), is given  by (see Eq.(\ref{cpeven}))
\begin{small}
\begin{eqnarray}{\label{MS}}
M^{2}_{b}\simeq\left(
\begin{array}{ccc}
M^{2}+m^{2} & 0 & kM\upsilon_{u}  \\
 *&   M^{2}+m^{2}+k^{2}\upsilon^{2}_{u}   & -kX_{k}\upsilon_{u} \\
* & * &  M^{2}+m^{2}+k^{2}\upsilon^{2}_{u} \\
\end{array}%
\right)~
\end{eqnarray}
\end{small}
where $X_{k}\simeq M+A_{k}-\mu\cot\beta$.
The neutral fermion mass squared matrix, which gives the mass eigenvalues $M^{2}_{f_{i}}$ in Eq.(\ref{potential}),  reads as (see also Eq.(\ref{neutralino}))
\begin{eqnarray}{\label{MF}}
M^{2}_{f}\simeq\left(
\begin{array}{ccc}
M^{2}+k^{2}\upsilon^{2}_{u} & 0 & kM\upsilon_{u}  \\
 *&   M^{2}  & kM\upsilon_{u} \\
* & * &  M^{2}+k^{2}\upsilon^{2}_{u} \\
\end{array}%
\right),
\end{eqnarray}

Substituting the mass eigenvalues in Eq.(\ref{MS}) and Eq.(\ref{MF}) into Eq.(\ref{potential}), 
one obtains the one-loop correction to SM-like Higgs mass
\begin{eqnarray}{\label{Higgsmassc}}
\Delta M^{2}_{h} \simeq \frac{1}{4\pi^{2}}k^{4}\upsilon^{2} \sin^{4}\beta\left[
\ln(x)+\frac{1}{8}\left(\frac{\lambda(y)}{x}-\lambda_{*}\right)- \frac{5}{192}\left(\frac{\lambda^{2}(y)}{x^{2}}-\lambda^{2}_{*}\right)\right]
\end{eqnarray}
where $x=1+m^{2}/M^{2}$, $y=X^{2}_{k}/M^{2}$ and $\lambda_{*}=\lambda(1)$, with $\lambda(y)$ defined as
\begin{eqnarray}{\label{gx}}
\lambda(y)=4+2y+2\sqrt{y^{2}+4y}.
\end{eqnarray}
As clearly seen in Eq.(\ref{Higgsmassc}),  $\Delta M^{2}_{h}\rightarrow 0$ under the degenerate mass limit $x\rightarrow 1$ and $y \rightarrow 1$, where the scalar and fermion masses are degenerate.
Conversely, large $\Delta M^{2}_{h}$ is expected in the situation with large $x$ and certain value of $y$.

\begin{figure}
\centering
\includegraphics[width=8cm,height=8cm]{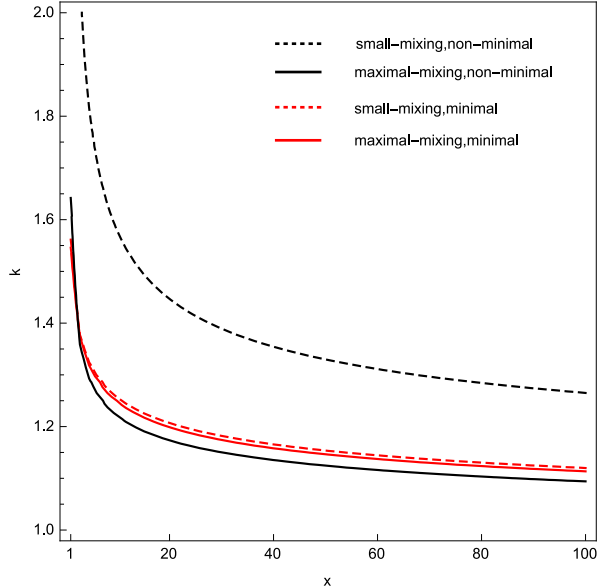}
\centering
 \caption{The contour of Higgs mass with $\Delta M_{h}=\sqrt{(125~\text{GeV})^2-M^{2}_{Z}}$ as function of Yukawa coupling constant  $k$ and ratio $x=1+m^{2}/M^{2}$ that describes the mass splitting between scalars and fermions in the VL electroweak sector.
 The red and black curves refer to the minimal and non-minimal VL model, respectively. }
\label{Higgs}
\end{figure}

Shown in Fig.\ref{Higgs} is the contour of Higgs mass with correction $\Delta M_{h}=\sqrt{(125~\text{GeV})^2-M^{2}_{Z}}$. 
The red curves represent the minimal VL model, which are compared with the 
non-minimal VL model \cite{0910.2732} as shown in black curves.
In individual case, the dashed and solid one refers to the small- and large-mixing effect, respectively.
There are a few comments in order.
$i)$, $X_{k}/M$, which is fixed by the condition of  maximal-mixing effect, 
is equal to $12x/5$ and $2(3x-1)$ in the minimal and non-minimal VL model, respectively. 
$ii)$, The discrepancies between the two models are small (large) in the situation with maximal (small)-mixing effect.
$iii)$, All of four patterns in Fig.\ref{Higgs} favor large value of $x$.
This suggests a large mass splitting ($m/M\sim 10$) between the scalars and fermions in the VL electroweak sector, 
if the Yukawa coupling $k$ is required to be in the perturbative region in high energy scales.
For VL fermions of order $\sim 1$ TeV, the VL scalars have mass of order $\sim 10$ TeV, 
which implies that the later ones have no role to play either at the LHC or dark matter experiments.

\section{Dark Matter}
In our model, the number of neutral fermions is seven,
 with four from the MSSM neutralino sector (for review, see, e.g.\cite{9506380}) 
and the other three from the VL sector.
The mass matrix for them is shown in Eq.(\ref{neutralino}),
where the complexity can be relatively reduced by taking the decoupling limit $v_{L}=\bar{v}_{L}=0$ as favored by precise measurements on the Higgs couplings.
Moreover, a small but nonzero $n$ is sufficient to yield desired decays in the VL electroweak sector,
with the first-order approximation to which 
the mass matrix in Eq.(\ref{neutralino}) is further divided into two separate parts - the neutralino mass matrix $A$ and the VL mass matrix $B$.

\subsection{VL Dark Matter}
For VL dark matter $\tilde{\chi}^{0}_{1}$ is decomposed as,
\begin{eqnarray}{\label{vld}}
\tilde{\chi}^{0}_{1}=N^{*}_{15}\tilde{\bar{\eta}}+N^{*}_{16}\tilde{\eta}+N^{*}_{17}\tilde{N},
\end{eqnarray}
where for simplicity we have neglected the neutralinos in Eq.(\ref{neutralino}).
The mass matrix $M_{\chi}$ is reduced to matrix $B$.
Since all of scalars in the VL sector are decoupled,
the Lagrangian in Eq.(\ref{VLs}) most relevant for this situation is given by\footnote{We will write both charged and neutral fermions in 4-component notation, 
e.g., $\chi^{+}_{E}=(\tilde{E}^{+}, \overline{\tilde{E}^{-}})$ and $\tilde{\chi}^{0}_{1}=(\chi^{0}_{1}, \overline{\chi^{0}_{1}})$.},
\begin{eqnarray}{\label{vllag}}
\mathcal{L}_{\text{VL}}&\supset& k(H^{+}_{u}\tilde{E}^{-}-H^{0}_{u}\tilde{\bar{\eta}})\tilde{N}-h(H^{0}_{d}\tilde{\eta}-H^{-}_{d}\tilde{E}^{+})\tilde{N}\nonumber\\
&-&\frac{g}{4c_{W}}\overline{\tilde{\eta}}\gamma^{\mu}\gamma_{5}\tilde{\eta} Z_{\mu}
+\frac{g}{4c_{W}}\overline{\tilde{\bar{\eta}}}\gamma^{\mu}\gamma_{5}\tilde{\bar{\eta}} Z_{\mu}\nonumber\\
&-&\frac{g}{\sqrt{2}}\overline{\tilde{E}^{+}}\gamma^{\mu}P_{L}\tilde{\eta}W^{+}_{\mu}+\frac{g}{\sqrt{2}}\overline{\tilde{E}^{-}}\gamma^{\mu}P_{R}\tilde{\bar{\eta}}W^{-}_{\mu}.
\end{eqnarray}
From Eq.(\ref{vllag}),
the VL $\tilde{\chi}^{0}_{1}$ annihilation is mediated by $Z$ boson, SM Higgs $h_{\text{SM}}$ and the charged fermion $\tilde{E}^{\pm}$, with SM final states $\bar{f}f$, $ZZ$, $WW$ and $h_{\text{SM}}h_{\text{SM}}$.

\begin{figure}
\centering
\includegraphics[width=8cm,height=8cm]{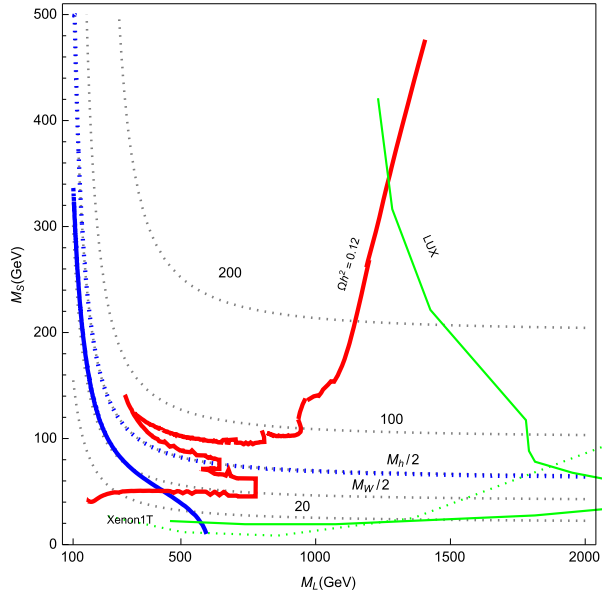}
\centering
 \caption{Constraints on the parameter space of VL dark matter with $k=1.2$ and $\tan\beta=20$. 
 We have shown the relic density of dark matter (red), the latest Xenon1T SI limit \cite{Xenon1T} (green dotted), LUX SD constraint \cite{LUXSD} (green solid), 
 $Z$ invisible decay limit \cite{Zdecay} (blue solid), and Higgs invisible decay limit \cite{hdecay} (blue dotted) simultaneously.
 The contours of VL dark matter masses in unit of GeV are shown in gray dotted curves.
 Regions on the top of green dotted curve, on the left side of green solid curve, and below the blue curves are excluded.}
\label{vldark}
\end{figure}

Eq.(\ref{vllag}) also indicates that
this dark matter model is a specific example of singlet-doublet dark matter \cite{0510064, 0705.4493, 0706.0918, 1109.2604, 1311.5896,1505.03867}. 
The key feature in our model is that the coupling $k\simeq 1.1-1.2$ due to the observed Higgs mass.
The other coupling $h$ can be neglected for $\tan\beta\geq 10$,
in which case we are left with two mass parameters $M_{L}$ and $M_{N}$.
Using these mass parameters, one obtains the dark matter couplings to Higgs and $Z$ \cite{1505.03867} given as respectively, 
\begin{eqnarray}{\label{effcoupling}}
c_{h}&\simeq&-\frac{k^{2}}{2}\frac{m_{\tilde{\chi}^{0}_{1}}\upsilon}{M^{2}_{L}+\frac{k^{2}}{2}\upsilon^{2}+2M_{N}m_{\tilde{\chi}^{0}_{1}}-3m^{2}_{\tilde{\chi}^{0}_{1}}},\nonumber\\
c_{Z}&\simeq&-\frac{k^{2}}{2} \frac{M_{Z}\upsilon(m^{2}_{\tilde{\chi}^{0}_{1}}-M^{2}_{L})}{(m^{2}_{\tilde{\chi}^{0}_{1}}-M^{2}_{L})^{2}+\frac{k^{2}}{2}\upsilon^{2}(m^{2}_{\tilde{\chi}^{0}_{1}}+M_{L}^{2})}.
\end{eqnarray}

Shown in Fig.\ref{vldark} are the constraints on the parameter space projected to the two-parameter plane of $M_{L}-M_{S}$ with $k=1.2$.
The red curve therein refers to the relic density of VL dark matter $\Omega h^2\simeq 0.12$.
We show the latest limits \cite{Xenon1T, LUXSD} on the spin-independent (SI) and spin-dependent (SD) dark matter-p/n scattering cross sections in green dotted and green solid plots, respectively,
where we have used the results $\sigma_{SI}\simeq c^{2}_{h}\times (2.11\times 10^{3})$ zb and 
$\sigma^{n}_{SD}\simeq c^{2}_{z}\times (8.97\times 10^{8})$ zb.
In order to be complete, indirect constraints are also shown in Fig.\ref{vldark}.
Since the VL leptons do not directly mix with the SM leptons, 
constraints such as lepton flavor violations are weaker than the electroweak precision measurements \cite{0803.4008} on the $Z$ boson and Higgs decay widths. 
In Fig.\ref{vldark}, the experimental bounds $\Delta \Gamma_{Z} \leq 2$ MeV \cite{Zdecay} and  $\Delta \Gamma_{h} \leq 0.16 \Gamma_{h}$ \cite{hdecay} are shown in blue solid and blue dotted curves, respectively, parameter regions below which are excluded.

The combination of these limits exclude parameter regions on the top of green dotted curve and on the left side of green solid curve.
It turns out that while VL dark matter with mass larger than $200$ GeV may be still consistent with the LUX SD limit, it is totally excluded by the latest Xenon1T limit.
The main reason for it is that large $k\sim1.2$ and large $\tan\beta$
lead to too large $c_h$ in Eq.(\ref{effcoupling}) to evade the Xenon1T limit
\footnote{Of course, this statement is not true in the situation where smaller $k$ can fit the observed Higgs mass when the one-loop correction to the Higgs mass is dominated by the stop parameters in the MSSM sector rather than the the electroweak fermions in the VL sector.}.
In the above analysis, we have handled the dark matter relic density by an analytic treatment similar to ref.\cite{1505.03867} instead of numerical calcualtions such as MicrOMEGAs \cite{1407.6129}.
In particular, we have used our previous results in ref.\cite{1711.10097}.

\subsection{Neutralino Dark Matter}
Although the VL dark matter is excluded, it still has a role to play in the case of neutralino dark matter.
A neutralino dark matter $\tilde{\chi}^{0}_{i}$ mainly arises from following components:
\begin{eqnarray}{\label{vld}}
\tilde{\chi}^{0}_{i}=N^{*}_{i1}\tilde{B}^{0}+N^{*}_{i2}\tilde{W}^{0}+N^{*}_{i3}\tilde{H}^{0}_{d}+N^{*}_{i4}\tilde{H}^{0}_{u},
\end{eqnarray}
whose relic density is affected by the VL sector in the following ways.

Firstly, $\tilde{\chi}^{0}_{1}$ pair may annihilate into the VL final states such as $\tilde{E^{+}}\tilde{E^{-}}$,
through either the $s$-channel exchange of $Z$ boson/Higgs scalar $h_i$ or the $t$- and $u$-channel exchange of VL scalars $E^{\pm}$.
However, this channel is not kinetically allowed,
since the masses of $\tilde{E^{\pm}}$ are larger than $m_{\tilde{\chi}^{0}_{1}}$ 
when $\tilde{E^{\pm}}$ share the same $R$-parity with the neutralino $\tilde{\chi}^{0}_{1}$.
Moreover, the $t$ and $u$-channel exchange of VL scalars $E^{\pm}$ are both suppressed by large VL scalar masses\footnote{If there are new sources to uplift the Higgs mass, the VL scalar masses can be of order sub-TeV, 
in which situation such process deserves a detailed study \cite{1608.00283,1810.07224}.}.

Second, the VL sector can mediate new Feynman diagrams for neutralino dark amtter annihilation:
\begin{itemize}
\item  Fermions $\tilde{E}^{\pm}$ mediate $t$- or $u$- channel Feynman diagrams for neutralino annihilation into $W^{\pm}$,
with the neutralino-$\chi^{\pm}_{E}$-W vertex given by
\begin{eqnarray}{\label{mediate1}}
\mathcal{L}\supset-\frac{g}{\sqrt{2}}W^{-}_{\mu}( N_{61}P_{L}-N_{51}P_{R})\overline{\tilde{\chi}^{0}_{1}} \gamma^{\mu}\chi^{+}_{E}+\text{H.c}.
\end{eqnarray}
\item Fermions $\tilde{\eta}$ and $\tilde{\bar{\eta}}$ mediate $t$- or $u$- channel Feynman diagrams for neutralino annihilation into $Z$ bosons,
with the neutralino-$\tilde{\eta}/\tilde{\bar{\eta}}$-Z vertex read as
\begin{eqnarray}{\label{mediate2}}
\mathcal{L}\supset -\frac{g}{4c_{W}}Z_{\mu}\overline{\tilde{\chi}^{0}_{1}}\gamma^{\mu}\gamma_{5}\left(N_{61}\tilde{\eta}-N_{51}\tilde{\bar{\eta}}\right)
\end{eqnarray}
\item Fermions $\tilde{\eta}$, $\tilde{\bar{\eta}}$ and $\tilde{N}$ mediate $t$- or $u$-channel Feynman diagrams for neutralino annihilation into $hh$,
with the neutralino-$\tilde{\eta}/\tilde{\bar{\eta}}/\tilde{N}$-$h_{\text{SM}}$ vertex in the MSSM modified by
\begin{eqnarray}{\label{mediate3}}
\mathcal{L}&\supset& 
\frac{1}{\sqrt{2}}(-kN_{51}s_{\beta}+hN_{61}c_{\beta})\overline{\tilde{\chi}^{0}_{1}}\tilde{N}h_{\text{SM}}
+\frac{1}{\sqrt{2}}N_{71}(-ks_{\beta}\tilde{\bar{\eta}}+hc_{\beta}\tilde{\eta})\overline{\tilde{\chi}^{0}_{1}}h_{\text{SM}}
\end{eqnarray}
\end{itemize}
Eq.(\ref{mediate1}) to Eq.(\ref{mediate3}) determine the deviation from ordinary neutralino,
which rely on mixings between $\tilde{\chi}^{1}_{0}$ and VL sector as described by components $N_{i1}$ with $i=5-7$.
With the mass parameter $m\sim$ several TeVs, 
the conditions of EWSB in Eq.(\ref{minimal}) imply that $\langle n\rangle$ is at most of order $\sim 10$ GeV and $\delta$ is less than $\sim 0.01$.
Thus, $N_{i1}$ reaches to its maximal values in the case of higgsino-like $\tilde{\chi}^{1}_{0}$.

\begin{figure}
\centering
\includegraphics[width=7cm,height=7cm]{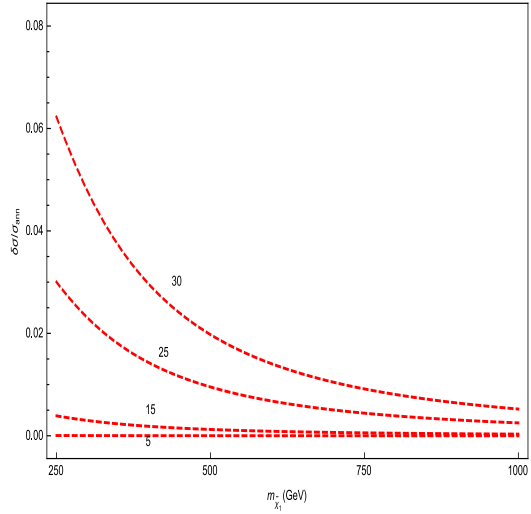}
\centering
 \caption{In the case of VL doublet-dominated correction, the ratio $\delta\sigma_{\text{ann}}/\sigma_{\text{ann}}$ as function of higgsino-like neutralino mass $m_{\tilde{\chi}^{0}_{1}}$ for $-h=k=1.2$, $\mid M_{L}-m_{\tilde{\chi}^{0}_{1}}\mid < M_{Z}$, 
 and different values of $n=\{5,15, 25, 30\}$ GeV, where $\sigma_{\text{ann}}\simeq 3\times 10^{-27}\text{cm}^{3}/s$.}
\label{neutralinoann}
\end{figure}

We divide the effects on higgsino-like $\tilde{\chi}^{1}_{0}$ into the VL doublets- and singlet-dominance.
In the case of VL doublet-dominated correction, $M_L$ is smaller than $M_{N}$.
The effect on the relic density of higgsino-like $\tilde{\chi}^{0}_{1}$ mainly arises from the neutralino-$\tilde{\eta}/\tilde{\bar{\eta}}$-Z vertex, 
which also contributes to the SD cross section. 
The first-order solution to the reduced $4\times 4$ mass matrix in Eq.(\ref{neutralino}) only composed of Higgs doublets and VL doublets is given by
\begin{eqnarray}{\label{ni1}}
N^{\tilde{H}_{d(u)}}_{61}&\simeq& -kn/(\mu+M_{L}), \nonumber\\
N^{\tilde{H}_{d(u)}}_{51}&\simeq& -(+) kn/(\mu+M_{L}), \nonumber\\
N^{\tilde{H}_{d(u)}}_{71}&\simeq& 0,
\end{eqnarray}
for $h= -k$ and $M_{N}>M_{L}$.

Substituting Eq.(\ref{ni1}) into Eq.(\ref{mediate2}), 
we show in Fig.\ref{neutralinoann} the maximal correction to dark matter annihilation cross section $\sigma_{\text{ann}}$ that occurs in the mass region 
$\mid M_{L}-m_{\tilde{\chi}^{0}_{1}}\mid < M_{Z}$ for $k=1.2$ and different values of $n=\{5, 15, 25, 30\}$ GeV.
In the mass range of $m_{\tilde{\chi}^{0}_{1}}=250-1000$ GeV, 
the ratio $\delta\sigma/\sigma_{\text{ann}}$ varies from $6\%$  ($n=30$ GeV) to less than $0.1\%$ ($n=5$ GeV).
It is sensitive to $n$ since $\delta\sigma$ is proportional to factor $(n/\mu+M_{L})^{4}$.
In the case of singlet-dominated correction, similar order of $\delta\sigma/\sigma_{\text{ann}}$ is  expected.

The modifications on the neutralino relic density, the SD and the SI cross sections, which are in percent level, suggest that 
the parameter space of neutralino dark matter is mainly expanded due to relaxing the Higgs mass constraint 
e.g. by lowering the values of $\tan\beta$ or gaugino masses in the case of universal soft masses.

\section{Collider Detection}
Collider detections on VL electroweak fermions depend on their interactions with SM particles.
If mixed with SM leptons such as $e$, $\mu$ or $\tau$,
VL  leptons can be constrained by mixing-relevant processes at collider constraints \cite{1506.01291,0107015, 1510.03456}.
These constraints depend on both the VL lepton mass and its mixing with the SM leptons.
In our case, the VL electroweak fermions $\chi_{E}^{\pm}$ ($\tilde{\eta}$, $\tilde{\bar{\eta}}$) directly mix with charginos (neutralinos) rather than SM leptons.
The strategies of searching them at colliders thus differ from the case of lepton mixings.

\begin{table}
\begin{center}
\begin{tabular}{ccc}
\hline\hline
Signals  &  Background   \\  \hline
$\chi_{E}^{+}\chi_{E}^{-}$ \cite{1403.5294}  &  jets + leptons +$E^{\rm miss}_{T}$ \\
$\tilde{\eta}$ or $\overline{\tilde{\eta}}$ pair \cite{1409.3168}    &  jets+ (leptons) +$E^{\rm miss}_{T}$ \\
$\chi_{E}^{\pm}$+$\tilde{\eta}$ or $\chi_{E}^{\pm}$+$\overline{\tilde{\eta}}$ \cite{1501.07110,1803.02762}   & (jets)+ leptons +$E^{\rm miss}_{T}$ \\
\hline \hline
\end{tabular}
\caption{The SM backgrounds for productions of VL electroweak fermion pairs at the LHC, 
where $E^{\rm miss}_{T}$ refers to missing transverse momentum.}
\label{states}
\end{center}
\end{table}

The VL electroweak fermions can be produced via chargino (neutralino) decay when their masses are beneath chargino (neutralino) masses.
Alternatively, they just imitate the electroweak productions of charginos (neutralinos).
The main difference between these two processes are that the cross section for the former case 
depends on both the VL lepton masses and their mixings with chargino or neutralinos,
while the cross section for the later one mainly relies on the VL electroweak fermion masses. 
Here, we stick to the later case, where the VL electroweak fermion pairs decay and are produced in terms of SM $Z$-, $W$- and Higgs-boson. 
Under the basis of gauge eigenstates as shown in Eq.(\ref{vllag}), 
the interaction vertexes of these processes are fixed by the SM gauge interactions,
so the numbers of events for VL lepton pairs are mainly sensitive to their mass parameters $M_L$ and $M_{N}$.
We show in Table.\ref{states} the backgrounds for the productions of various VL electroweak fermion paris at the LHC,
where mass bounds on neutralinos and charginos can be found.
$m_{\tilde{\chi}^{\pm}_{1}}$ or $m_{\tilde{\chi}^{0}_{2}}$ up to $\sim 550$ GeV is excluded for $m_{\tilde{\chi}^{0}_{1}}$ less than $\sim 100$ GeV,
and they are obviously relaxed when $m_{\tilde{\chi}^{0}_{1}}$ is above $100$ GeV \cite{1501.07110,1803.02762}.

The mixing effects in the VL electroweak sector are described by the mass matrix $B$ in Eq.(\ref{ABC}),
which is similar to the higgsino-bino benchmark scenario. 
See the reduced matrix A with wino decoupled in Eq.(\ref{ABC}).
We can infer the implications of LHC limits above to the parameter space of $M_{L}-M_{N}$ by introducing a ratio of signal strengths $\hat{\mu}$.
For example, for the process $pp\rightarrow \tilde{\chi}^{\pm}_{E}\tilde{\chi}^{0}_{2}+$ jets $+$ leptons \cite{1501.07110,1803.02762} we have
\begin{eqnarray}{\label{scaling}}
\hat{\mu}=\frac{\sigma(pp\rightarrow \tilde{\chi}^{\pm}_{E}\tilde{\chi}^{0}_{2}+\cdots)}{\sigma(pp\rightarrow \tilde{\chi}^{\pm}_{1}\tilde{\chi}^{0}_{2}+\cdots)}
\cdot
\frac{\text{Br}(\tilde{\chi}^{\pm}_{E}\rightarrow W^{\pm} \tilde{\chi}^{0}_{1})}{\text{Br}(\tilde{\chi}^{\pm}_{1}\rightarrow W^{\pm} \tilde{\chi}^{0}_{1})}\cdot
\frac{\text{Br}(\tilde{\chi}^{0}_{2}\rightarrow Z \tilde{\chi}^{0}_{1})}{\text{Br}(\tilde{\chi}^{0}_{2}\rightarrow Z \tilde{\chi}^{0}_{1})},
\end{eqnarray}
where $\tilde{\chi}^{0}_{1}$ in the numerators (denominators) refers to the lightest neutral fermion in the VL (higgsino-bino) sector.
In the parameter regions with $M_{N}<M_{L}$, where $\tilde{\chi}^{0}_{1}\sim \tilde{N}^{0}$, $\tilde{\chi}^{0}_{2}\sim \tilde{\eta}^{0}$
and $\tilde{\chi}^{\pm}_{E}\sim \tilde{E}^{\pm}$, 
the number of events for $\tilde{\chi}^{\pm}_{E}\tilde{\chi}^{0}_{2}$ pair can be of same order as that of $\tilde{H}_{u,d}^{\pm}\tilde{B}^{0}$.
So the pattern of LHC limits projected to the plane of $M_{N}-M_{L}$ is similar to Fig.8(d) in ref.\cite{1803.02762}.
In the parameter regions with $M_{N}>M_{L}$, where $\tilde{\chi}^{0}_{1}\sim \tilde{\eta}^{0}$, $\tilde{\chi}^{0}_{2}\sim \tilde{N}^{0}$,
and $\tilde{\chi}^{\pm}_{E}\sim \tilde{E}^{\pm}$, the mass splitting between $ \tilde{E}^{\pm}$ and $\tilde{\chi}^{0}_{1}$ is small similar to the case of Higgsino-like  $\tilde{\chi}^{\pm}_{1}$ and $\tilde{\chi}^{0}_{1}$.
In this situation, the decays  \cite{0104115} of charged fermion $\tilde{\chi}^{\pm}_{E}\rightarrow \tilde{\chi}^{0}_{1}W^{*\pm}\rightarrow\tilde{\chi}^{0}_{1} \cdots$ are rather sensitive to the mass splitting.

\section{Conclusion}
In this paper, we have studied the minimal VL model motivated by the grand unification.
The key feature in such model is that the VL electroweak sector couples to the Higgs doublets rather than mixes with the SM leptons.
Therefore, they can play important roles both in the phenomenologies of Higgs physics and dark matter. 

For the Higgs physics, we have used the LHC bounds on the Higgs coupling constants to constrain the vevs in the VL electroweak sector,
the magnitude of which are found to be small.
Consequently, the magnitudes of the mixing effects between the Higgs doublets and the VL electroweak sector controlled by the vevs are small as well,
leading to small deviation in the Higgs self coupling from the MSSM expectation.
Moreover, we used the observed Higgs mass to constrain the other model parameters in the VL sector.
The fit reveals that both large Yukawa coupling constant $k\sim1.1$ and large mass splitting between scalar and fermion masses in the VL sector are required.

For the dark matter phenomenology,
w have verified that the large Yukawa coupling $k$ excludes the possibility of VL dark matter.
For it gives rise to too large dark matter coupling to Higgs to evade the Xenon1T limit.
Nevertheless, they are still useful to reduce the tension on the neutralino dark matter by relaxing the constraint from Higgs mass.

Finally, in order to be complete we briefly discussed the constraints on the VL electroweak  fermions at the LHC.
Unlike in the 4-th generation leptons which directly mix with SM leptons,
the VL  electroweak fermions can be either produced via the decays of charginos and neutralinos,
or they imitate the electroweak productions of charginos and neutralinos.
The prospect of the later class at the high luminosity-LHC will be explored elsewhere \cite{Xu}.

\begin{acknowledgments}
We would like to thank P. Fayet for correspondence and H. Wu for a numerical check in the case of neutralino dark matter. 
This research is supported by the National Natural Science Foundation of China under Grant No.11775039,
the Chinese Scholarship Council, 
and the Fundamental Research Funds for the Central Universities at Chongqing University under Grant No.cqu2017hbrc1B05.
\end{acknowledgments}

\appendix
\section{Scalar Mass Matrix}
\subsection{CP-even Scalar}
In the basis $\phi_{S}^{T}=(H^{0}_{uR}, H^{0}_{dR},\eta_{R}, \bar{\eta}_{R}, N_{R})$,
the matrix elements for symmetric mass matrix squared $\mathcal{M}_{S}^{2}$ of CP-even scalars in the Lagrangian 
\begin{eqnarray}{\label{cpeven}}
\mathcal{L}\supset \frac{1}{2}\phi_{S}^{T,i}\mathcal{M}^{2}_{S,ij}\phi_{S}^{j}
\end{eqnarray}
are as follows,
\begin{eqnarray}{\label{cpevenc}}
\mathcal{M}^{2}_{S,11}&=&\mu^{2}+m^{2}_{H_{u}}+k^{2}(n^{2}+\bar{v}^{2}_{L})\nonumber\\
\mathcal{M}^{2}_{S,12}&=&b-khv_{L}\bar{v}_{L}\nonumber\\
\mathcal{M}^{2}_{S,13}&=&-hn\mu-kh \bar{v}_{L}\upsilon_{d} +k M_{L}n \nonumber\\
\mathcal{M}^{2}_{S,14}&=&2k^{2}\upsilon_{u}\bar{v}_{L}+k(M_{N}n+h\upsilon_{d}v_{L})-kA_{k}n\nonumber\\
\mathcal{M}^{2}_{S,15}&=&2k^{2}n\upsilon_{u}-hv_{L}\mu-kM_{N}\bar{v}_{L}+kM_{L}v_{L}-kA_{k}\bar{v}_{L} \nonumber\\
\mathcal{M}^{2}_{S,22}&=&\mu^{2}+ m^{2}_{H_{d}} +h^{2}(n^{2}+v^{2}_{L}) \nonumber\\
\mathcal{M}^{2}_{S,23}&=&2h^{2}\upsilon_{d}v_{L}+h(M_{N}n-k\upsilon_{u}\bar{v}_{L})-hA_{h}n \nonumber\\
\mathcal{M}^{2}_{S,24}&=&kn\mu -kh\upsilon_{u}v_{L}-hM_{L}n\nonumber\\
\mathcal{M}^{2}_{S,25}&=& 2h^{2}n\upsilon_{d}+k\bar{v}_{L}\mu+hM_{N}v_{L}+hM_{L}\bar{v}_{L}+hA_{k}v_{L} \nonumber\\
\mathcal{M}^{2}_{S,33}&=&M^{2}_{L}+m^{2}_{L}+h^{2}(n^{2}+\upsilon_{d}^{2})\nonumber\\
\mathcal{M}^{2}_{S,34}&=&-kh \upsilon_{u}\upsilon_{d} \nonumber\\
\mathcal{M}^{2}_{S,35}&=&2h^{2}nv_{L}-h\upsilon_{u}\mu+hM_{N}\upsilon_{d}+kM_{L}\upsilon_{u}+hA_{h}\upsilon_{d}  \nonumber\\
\mathcal{M}^{2}_{S,44}&=&M^{2}_{L}+m^{2}_{\bar{L}}+k^{2}(n^{2}+\upsilon^{2}_{u})\nonumber\\
\mathcal{M}^{2}_{S,45}&=&2k^{2}n\bar{v}_{L}+k\upsilon_{d}\mu-kM_{N}\upsilon_{u}-hM_{L}\upsilon_{d}-kA_{k}\upsilon_{u}\nonumber\\
\mathcal{M}^{2}_{S,55}&=& M^{2}_{N}+m^{2}_{N}+h^{2}(v^{2}_{L}+\upsilon^{2}_{d})+k^{2}(\bar{v}^{2}_{L}+\upsilon^{2}_{u})
\end{eqnarray}
where $D$-term contributions have been neglected. 
After diagonalizing the matrix $\mathcal{M}_{S}^{2}$, 
we obtain five physical neutral scalars, one of which severs as the SM-like Higgs boson $h$ with mass $125$ GeV \cite{Higgsmass1, Higgsmass2}. 

\subsection{CP-odd Scalar}
In the basis $\phi_{A}^{T}=(H_{uI},H_{dI},\eta_{I},\bar{\eta}_{I}, N_{I})$ the matrix elements for CP-odd scalars in the Lagrangian 
\begin{eqnarray}{\label{cpodd}}
\mathcal{L}\supset \frac{1}{2}\phi_{A}^{T,i}\mathcal{M}^{2}_{A,ij}\phi^{j}_{A}
\end{eqnarray}
are given by,
\begin{eqnarray}{\label{cpoddc}}
\mathcal{M}^{2}_{A,11}&=&\mu^{2}+m^{2}_{H_{u}}+k^{2}(n^{2}+\bar{v}^{2}_{L}) \nonumber\\
\mathcal{M}^{2}_{A,12}&=&b-khv_{L}\bar{v}_{L}\nonumber\\
\mathcal{M}^{2}_{A,13}&=&-hn\mu-kh \bar{v}_{L}\upsilon_{d} +k M_{L}n\nonumber\\
\mathcal{M}^{2}_{A,14}&=&-k(M_{N}n+h\upsilon_{d}v_{L})+kA_{k}n\nonumber\\
\mathcal{M}^{2}_{A,15}&=&-hv_{L}\mu-kM_{N}\bar{v}_{L}-kM_{L}v_{L}+kA_{k}\bar{v}_{L} \nonumber\\
\mathcal{M}^{2}_{A,22}&=&\mu^{2}+m^{2}_{H_{d}} +h^{2}(n^{2}+v^{2}_{L}) \nonumber\\
\mathcal{M}^{2}_{A,23}&=&-h(M_{N}n-k\upsilon_{u}\bar{v}_{L})-hA_{h}n \nonumber\\
\mathcal{M}^{2}_{A,24}&=&kn\mu -kh\upsilon_{u}v_{L}-hM_{L}n\nonumber\\
\mathcal{M}^{2}_{A,25}&=&+k\bar{v}_{L}\mu+hM_{N}v_{L}-hM_{L}\bar{v}_{L}-hA_{k}v_{L} \nonumber\\
\mathcal{M}^{2}_{A,33}&=&M^{2}_{L}+m^{2}_{L}+h^{2}(n^{2}+\upsilon_{d}^{2})\nonumber\\
\mathcal{M}^{2}_{A,34}&=&-kh \upsilon_{u}\upsilon_{d}\nonumber\\
\mathcal{M}^{2}_{A,35}&=&h\upsilon_{u}\mu+hM_{N}\upsilon_{d}+kM_{L}\upsilon_{u}-hA_{h}\upsilon_{d} \nonumber\\
\mathcal{M}^{2}_{A,44}&=&M^{2}_{L}+m^{2}_{\bar{L}}+k^{2}(n^{2}+\upsilon^{2}_{u})\nonumber\\
\mathcal{M}^{2}_{A,45}&=&-k\upsilon_{d}\mu-kM_{N}\upsilon_{u}-hM_{L}\upsilon_{d}+kA_{k}\upsilon_{u}\nonumber\\
\mathcal{M}^{2}_{A,55}&=& M^{2}_{N}+m^{2}_{N}+h^{2}(v^{2}_{L}+\upsilon^{2}_{d})+k^{2}(\bar{v}^{2}_{L}+\upsilon^{2}_{u})\nonumber\\
\end{eqnarray}
where the $D$-term contributions have been neglected.
Under the decoupling limit $v_{L}=\bar{v}_{L}=0$ and $n=0$, scalar $\eta_{I}$, $\bar{\eta}_{I}$ and $N_{I}$ decouple from $H_{uI}$ and $H_{dI}$, 
which results in the well known result $\text{Det}\mathcal{M}^{2}_{A} =0$.

\subsection{CP-charged Scalar}
In the basis $\phi_{c}^{T}=(H^{+}_{u},H^{-*}_{d}, E^{+}, E^{-*})$ the matrix elements for charged scalars in the Lagrangian 
\begin{eqnarray}{\label{cpodd}}
\mathcal{L}\supset \phi_{c}^{* T, i}\mathcal{M}^{2}_{C,ij}\phi^{j}_{c}
\end{eqnarray}
read as 
\begin{eqnarray}{\label{cpoddc}}
\mathcal{M}^{2}_{C,11}&=&\mu^{2}+m^{2}_{H_{u}}+k^{2}(n^{2}+\bar{v}^{2}_{L}) \nonumber\\
\mathcal{M}^{2}_{C,12}&=&-khv_{L}\bar{v}_{L}\nonumber\\
\mathcal{M}^{2}_{C,13}&=&-hn\mu-kh \bar{v}_{L}\upsilon_{d} +k M_{L}n\nonumber\\
\mathcal{M}^{2}_{C,14}&=&k(M_{N}n-h\upsilon_{d}v_{L})+kA_{k}n\nonumber\\
\mathcal{M}^{2}_{C,22}&=&\mu^{2}+ m^{2}_{H_{d}} +h^{2}(n^{2}+v^{2}_{L}) \nonumber\\
\mathcal{M}^{2}_{C,23}&=&-h(M_{N}n-k\upsilon_{u}\bar{v}_{L})-hA_{h}n \nonumber\\
\mathcal{M}^{2}_{C,24}&=& kn\mu-kh\upsilon_{u}v_{L}-hM_{L}n\nonumber\\
\mathcal{M}^{2}_{C,33}&=&M^{2}_{L}+m^{2}_{L}+h^{2}(n^{2}+\upsilon_{d}^{2})\nonumber\\
\mathcal{M}^{2}_{C,34}&=&-kh \upsilon_{u}\upsilon_{d}\nonumber\\
\mathcal{M}^{2}_{C,44}&=&M^{2}_{L}+m^{2}_{\bar{L}}+k^{2}(n^{2}+\upsilon^{2}_{u})
\end{eqnarray}
Under the decoupling limit $v_{L}=\bar{v}_{L}=0$ and $n=0$, scalar $E^{+}$ and $E^{-*}$ decouple from the others, 
due to which one recovers the relation $\text{Det}\mathcal{M}^{2}_{C} =0$.\\

\section{Fermion Mass Matrix}

\subsection{Charged Fermions}
In the basis $\chi^{T}_{\pm}=(\tilde{W^{+}}, \tilde{H_{u}^{+}}, \tilde{E^{+}}, \tilde{W^{-}}, \tilde{H^{-}_{d}}, \tilde{E^{-}})$ 
the mass matrix in $\mathcal{L}\supset -\frac{1}{2} \chi_{\pm}^{T, i}\mathcal{M}_{\chi^{\pm},ij}\chi_{\pm}^{ j}$
is given by,
\begin{eqnarray}{\label{chargino2}}
\mathcal{M}_{\chi^{\pm},ij}=\left(
\begin{array}{cc}
0& X^{T}  \\
X & 0 \\
\end{array}%
\right)
\end{eqnarray}
with
\begin{small}
\begin{eqnarray}{\label{X}}
X=\left(
\begin{array}{ccc}
M_{2}& \sqrt{2(1-\delta)}s_{\beta}M_{W}  &  \sqrt{2\delta}s_{\gamma}M_{W}   \\
\sqrt{2(1-\delta)}c_{\beta}M_{W}  &  \mu  & -hn  \\
\sqrt{2\delta}c_{\gamma}M_{W}  &  kn & M_{L}\\
\end{array}%
\right)\nonumber\\
\end{eqnarray}
\end{small}
where $s_{\beta}=\sin\beta$, $c_{\beta}=\cos\beta$, $s_{\gamma}=\sin\gamma$ and $c_{\gamma}=\cos\gamma$.

\subsection{Neutral Fermions}
In the basis $\psi_{0}^{T}=(\tilde{B}^{0}, \tilde{W}^{0}, \tilde{H^{0}_{d}}, \tilde{H^{0}_{u}}, \tilde{\bar{\eta}},  \tilde{\eta},\tilde{N})$ 
the mass matrix  in $\mathcal{L}\supset -\frac{1}{2} \psi_{0}^{T, i}\mathcal{M}_{\chi,ij}\psi_{0}^{ j}$ reads as,
\begin{eqnarray}{\label{neutralino}}
M_{\chi}=\left(
\begin{array}{cc}
A& C\\
 *& B \\
\end{array}%
\right),
\end{eqnarray}
with
\begin{small}
\begin{eqnarray}{\label{ABC}}
A&=&\left(
\begin{array}{cccc}
M_{1} & 0 &  -(1-\delta)M_{Z}s_{W}c_{\beta}&  (1-\delta)M_{Z}s_{W}s_{\beta} \\
 *&   M_{2}  & (1-\delta)M_{Z}c_{W}c_{\beta}  &  -(1-\delta)M_{Z}c_{W}s_{\beta}  \\
* & * & 0 & -\mu \\
* & * & * & 0 
\end{array}%
\right),\nonumber\\
B&=&\left(
\begin{array}{ccc}
0& -M_{L} & -k\upsilon_{u} \\
 * &0  &h\upsilon_{d} \\
  * &* & -M_{N}\\
\end{array}%
\right),\nonumber\\
C&=&\left(
\begin{array}{cccc}
 -\delta M_{Z}s_{W}c_{\gamma} & \delta M_{Z}s_{W}s_{\gamma} & 0\\
 \delta M_{Z}c_{W}c_{\gamma} & -\delta M_{Z}c_{W}s_{\gamma} & 0 \\
0 &  hn  & hv_{L} \\
-kn   & 0& -k\bar{v}_{L} \\
\end{array}%
\right).
\end{eqnarray}
\end{small}


\begin{thebibliography}{99}
 \bibitem{Higgsmass1}
G.\ Aad {\it et al}. [ATLAS Collaboration], 
Phys.\ Lett.\ B{\bf716}, 1 (2012),  
arXiv:1207.7214 [hep-ex].

\bibitem{Higgsmass2}
S.\ Chatrchyan {\it et al}. [CMS Collaboration], 
Phys.\ Lett.\ B{\bf716}, 30 (2012),
arXiv:1207.7235 [hep-ex].

\bibitem{MSSMHiggs1}
 M.~Carena, S.~Gori, N.~R.~Shah and C.~E.~M.~Wagner,
JHEP {\bf 1203}, 014 (2012),
[arXiv:1112.3336 [hep-ph]].

\bibitem{MSSMHiggs2}
 L.~J.~Hall, D.~Pinner and J.~T.~Ruderman,
JHEP {\bf 1204}, 131 (2012),
[arXiv:1112.2703 [hep-ph]].


\bibitem{MSSMDark1}
H.~Baer, V.~Barger and H.~Serce,
Phys.\ Rev.\ D {\bf 94}, no. 11, 115019 (2016),
[arXiv:1609.06735 [hep-ph]].

\bibitem{MSSMDark2}
 J.~Cao, Y.~He, L.~Shang, W.~Su, P.~Wu and Y.~Zhang,
JHEP {\bf 1610}, 136 (2016),
[arXiv:1609.00204 [hep-ph]].



\bibitem{1711.05362}
  S.~Zheng,
Phys.\ Rev.\ D {\bf 98}, 035028 (2018),
[arXiv:1711.05362 [hep-ph]].

\bibitem{1706.01071}
 S.~Zheng,
Eur.\ Phys.\ J.\ C {\bf 77}, no. 9, 588 (2017),
[arXiv:1706.01071 [hep-ph]].

\bibitem{0910.2732}
  S.~P.~Martin,
Phys.\ Rev.\ D {\bf 81}, 035004 (2010),
[arXiv:0910.2732 [hep-ph]].

\bibitem{MO}
 T.~Moroi and Y.~Okada,
  Phys.\ Lett.\ B {\bf 295}, 73 (1992).

\bibitem{Babu}
K.~S.~Babu, I.~Gogoladze, M.~U.~Rehman and Q.~Shafi,
Phys.\ Rev.\ D {\bf 78}, 055017 (2008).
  

\bibitem{Fayet}
P.~Fayet,
Nucl.\ Phys.\ B {\bf 90}, 104 (1975).


\bibitem{9709356}
  S.~P.~Martin,
Adv.\ Ser.\ Direct.\ High Energy Phys.\  {\bf 21}, 1 (2010),
[hep-ph/9709356].

\bibitem{1709.07497}
A.~M.~Sirunyan {\it et al.} [CMS Collaboration],
Phys.\ Lett.\ B {\bf 780}, 501 (2018),
[arXiv:1709.07497 [hep-ex]].


\bibitem{1501.04943}
 G.~Aad {\it et al.} [ATLAS Collaboration],
JHEP {\bf 1504}, 117 (2015),
[arXiv:1501.04943 [hep-ex]].

\bibitem{1412.2641}
G.~Aad {\it et al.} [ATLAS Collaboration],
Phys.\ Rev.\ D {\bf 92}, no. 1, 012006 (2015),
[arXiv:1412.2641 [hep-ex]].

\bibitem{1303.3879}
  J.~Ellis and T.~You,
JHEP {\bf 1306}, 103 (2013),
[arXiv:1303.3879 [hep-ph]].

\bibitem{1606.02266}
 G.~Aad {\it et al.} [ATLAS and CMS Collaborations],
JHEP {\bf 1608}, 045 (2016),
[arXiv:1606.02266 [hep-ex]].

\bibitem{1809.10733}
 A.~M.~Sirunyan {\it et al.} [CMS Collaboration],
Eur.\ Phys.\ J.\ C {\bf 79}, no. 5, 421 (2019),
[arXiv:1809.10733 [hep-ex]].

\bibitem{1310.8361}
S.~Dawson {\it et al.},
[arXiv:1310.8361 [hep-ex]].

\bibitem{1310.0763}
 D.~M.~Asner {\it et al.},
[arXiv:1310.0763 [hep-ph]].

\bibitem{1412.8662}
  V.~Khachatryan {\it et al.} [CMS Collaboration],
Eur.\ Phys.\ J.\ C {\bf 75}, no. 5, 212 (2015),
[arXiv:1412.8662 [hep-ex]].

\bibitem{CW}
S.~R.~Coleman and E.~J.~Weinberg,
Phys.\ Rev.\ D {\bf 7}, 1888 (1973).


\bibitem{9603205}
 T.~Plehn, M.~Spira and P.~M.~Zerwas,
Nucl.\ Phys.\ B {\bf 479}, 46 (1996), Erratum: [Nucl.\ Phys.\ B {\bf 531}, 655 (1998)],
[hep-ph/9603205].

\bibitem{GB}
E.~W.~N.~Glover and J.~J.~van der Bij,
Nucl.\ Phys.\ B {\bf 309}, 282 (1988).

\bibitem{9805244}
S.~Dawson, S.~Dittmaier and M.~Spira,
Phys.\ Rev.\ D {\bf 58}, 115012 (1998),
[hep-ph/9805244].

\bibitem{1305.7340}
 J.~Grigo, J.~Hoff, K.~Melnikov and M.~Steinhauser,
Nucl.\ Phys.\ B {\bf 875}, 1 (2013),
[arXiv:1305.7340 [hep-ph]].

\bibitem{1309.6594}
D.~de Florian and J.~Mazzitelli,
Phys.\ Rev.\ Lett.\  {\bf 111}, 201801 (2013),
[arXiv:1309.6594 [hep-ph]].

\bibitem{1206.5816}
N.~D.~Christensen, T.~Han and T.~Li,
Phys.\ Rev.\ D {\bf 86}, 074003 (2012),
[arXiv:1206.5816 [hep-ph]].

\bibitem{1304.3670}
  R.~Barbieri, D.~Buttazzo, K.~Kannike, F.~Sala and A.~Tesi,
Phys.\ Rev.\ D {\bf 87}, no. 11, 115018 (2013),
[arXiv:1304.3670 [hep-ph]].

\bibitem{1305.6397}
  R.~S.~Gupta, H.~Rzehak and J.~D.~Wells,
Phys.\ Rev.\ D {\bf 88}, 055024 (2013),
[arXiv:1305.6397 [hep-ph]].

\bibitem{1504.06932}
L.~Wu, J.~M.~Yang, C.~P.~Yuan and M.~Zhang,
Phys.\ Lett.\ B {\bf 747}, 378 (2015),
[arXiv:1504.06932 [hep-ph]].


\bibitem{9506380}
G.~Jungman, M.~Kamionkowski and K.~Griest,
Phys.\ Rept.\  {\bf 267}, 195 (1996),
[hep-ph/9506380].
 

\bibitem{0510064}
  R.~Mahbubani and L.~Senatore,
Phys.\ Rev.\ D {\bf 73}, 043510 (2006),
[hep-ph/0510064].

\bibitem{0705.4493}
  F.~D'Eramo,
Phys.\ Rev.\ D {\bf 76}, 083522 (2007),
[arXiv:0705.4493 [hep-ph]].

\bibitem{0706.0918}
  R.~Enberg, P.~J.~Fox, L.~J.~Hall, A.~Y.~Papaioannou and M.~Papucci,
JHEP {\bf 0711}, 014 (2007),
[arXiv:0706.0918 [hep-ph]].


\bibitem{1109.2604}
 T.~Cohen, J.~Kearney, A.~Pierce and D.~Tucker-Smith,
Phys.\ Rev.\ D {\bf 85}, 075003 (2012),
[arXiv:1109.2604 [hep-ph]].

\bibitem{1311.5896}
  C.~Cheung and D.~Sanford,
JCAP {\bf 1402}, 011 (2014),
[arXiv:1311.5896 [hep-ph]].

\bibitem{1505.03867}
 L.~Calibbi, A.~Mariotti and P.~Tziveloglou,
JHEP {\bf 1510}, 116 (2015),
[arXiv:1505.03867 [hep-ph]].



\bibitem{Xenon1T}
  E.~Aprile {\it et al.} [XENON Collaboration],
Phys.\ Rev.\ Lett.\  {\bf 121}, no. 11, 111302 (2018),
[arXiv:1805.12562 [astro-ph.CO]].

\bibitem{LUXSD}
D.~S.~Akerib {\it et al.} [LUX Collaboration],
Phys.\ Rev.\ Lett.\  {\bf 118}, no. 25, 251302 (2017),
[arXiv:1705.03380 [astro-ph.CO]].


\bibitem{0803.4008}
 F.~del Aguila, J.~de Blas and M.~Perez-Victoria,
Phys.\ Rev.\ D {\bf 78}, 013010 (2008),
[arXiv:0803.4008 [hep-ph]].

\bibitem{Zdecay}
 S.~Schael {\it et al.} [ALEPH and DELPHI and L3 and OPAL and SLD Collaborations and LEP Electroweak Working Group and SLD Electroweak Group and SLD Heavy Flavour Group],
Phys.\ Rept.\  {\bf 427}, 257 (2006),
[hep-ex/0509008].

\bibitem{hdecay}
 G.~Aad {\it et al.} [ATLAS and CMS Collaborations],
JHEP {\bf 1608}, 045 (2016),
[arXiv:1606.02266 [hep-ex]].



\bibitem{1407.6129}
G.~Belanger, F.~Boudjema, A.~Pukhov and A.~Semenov,
 Comput.\ Phys.\ Commun.\  {\bf 192}, 322 (2015),
 arXiv:1407.6129 [hep-ph].

\bibitem{1711.10097}
  H.~Han, H.~Wu and S.~Zheng,
Chin.\ Phys.\ C {\bf 43}, 043103 (2019),
[arXiv:1711.10097 [hep-ph]].


\bibitem{1608.00283}
  M.~Abdullah, J.~L.~Feng, S.~Iwamoto and B.~Lillard,
Phys.\ Rev.\ D {\bf 94}, no. 9, 095018 (2016),
[arXiv:1608.00283 [hep-ph]].

\bibitem{1810.07224}
J.~Y.~Araz, S.~Banerjee, M.~Frank, B.~Fuks, A.~Goudelis, 
 arXiv:1810.07224 [hep-ph].
 
 
 

\bibitem{1506.01291}
G.~Aad {\it et al.} [ATLAS Collaboration],
JHEP {\bf 1509}, 108 (2015),
[arXiv:1506.01291 [hep-ex]].

\bibitem{0107015}
 P.~Achard {\it et al.} [L3 Collaboration],
Phys.\ Lett.\ B {\bf 517}, 75 (2001),
[hep-ex/0107015].

\bibitem{1510.03456}
 N.~Kumar and S.~P.~Martin,
Phys.\ Rev.\ D {\bf 92}, no. 11, 115018 (2015),
[arXiv:1510.03456 [hep-ph]].


\bibitem{1403.5294}
G.~Aad {\it et al.} [ATLAS Collaboration],
JHEP {\bf 1405}, 071 (2014),
[arXiv:1403.5294 [hep-ex]].

\bibitem{1409.3168}
 V.~Khachatryan {\it et al.} [CMS Collaboration],
Phys.\ Rev.\ D {\bf 90}, no. 9, 092007 (2014),
[arXiv:1409.3168 [hep-ex]].

\bibitem{1501.07110}
G.~Aad {\it et al.} [ATLAS Collaboration],
Eur.\ Phys.\ J.\ C {\bf 75}, no. 5, 208 (2015),
[arXiv:1501.07110 [hep-ex]].

\bibitem{1803.02762}
 M.~Aaboud {\it et al.} [ATLAS Collaboration],
Eur.\ Phys.\ J.\ C {\bf 78}, no. 12, 995 (2018),
[arXiv:1803.02762 [hep-ex]].

\bibitem{0104115}
 A.~Djouadi, Y.~Mambrini and M.~Muhlleitner,
Eur.\ Phys.\ J.\ C {\bf 20}, 563 (2001),
[hep-ph/0104115].

\bibitem{Xu}
 S.~Xu and S.~Zheng,
[arXiv:1912.00404 [hep-ph]].

\end{thebibliography}
\end{document}